%% file: ms.tex
\documentclass[]{raa}            
\usepackage{graphicx,times}
\usepackage{natbib}

\providecommand{\bm}[1]{\mbox{\boldmath$#1$\unboldmath}}
\def\kms{$\rm{km~s}^{-1}$}
\def\etal{et al.}

\begin{document}

   \title{Multicolor Photometry of the Galaxy Cluster A98:
   Substructures and Star Formation Properties}

 \volnopage{ {\bf 2009} Vol.\ {\bf 9} No. {\bf XX}, 000--000}
   \setcounter{page}{1}

   \author{L. Zhang \inst{1}
   \and Q.-R. Yuan  \inst{1}
   \and X. Zhou \inst{2}
   \and Z.-J. Jiang \inst{2}
   \and Y.-B. Yang \inst{2}
   \and J. Ma \inst{2}
   \and J.-H. Wu \inst{2}
   \and Z.-Y. Wu \inst{2}
   }

   \institute{Department of Physics, Nanjing Normal University,
                 Wenyuan Road 3, Nanjing 210046, China; \\
   \hspace{3.5mm}{Email: \it yuanqirong@njnu.edu.cn; lizhang722@163.com} \\
        \and
             National Astronomical Observatories, Chinese Academy of Sciences,
             Beijing 100012, China.\\
        \vs \no
}

\abstract{An optical photometric observation with the
Beijing-Arizona-Taiwan-Connecticut (BATC) multicolor system is
carried out for A98 ($z=0.104$), a galaxy cluster with two large
enhancements in X-ray surface brightness. The spectral energy
distributions (SEDs) covering 15 intermediate bands are obtained
for all sources detected down to $V \sim 20$ mag in a field of
$58' \times 58'$. After the star-galaxy separation by the
color-color diagrams, a photometric redshift technique is applied
to the galaxy sample for further membership determination. The
color-magnitude relation is taken as a further restriction of the
early-type cluster galaxies. As a result, a list of 198 faint
member galaxies is achieved. Based on newly generated sample of
member galaxies, the dynamical substructures, A98N, A98S, and
A98W, are investigated in detail. A separate galaxy group, A98X,
is also found to the south of main concentration of A98, which is
gravitationally unbound to A98.
For 74 spectroscopically confirmed member galaxies, the
environmental effect on the star formation histories is found. The
bright galaxies in the core region are found to have shorter time
scales of star formation, longer mean stellar ages, and higher
metallicities of interstellar medium, which can be interpreted in
the context of hierarchical cosmological scenario.
 }

   \authorrunning{L. Zhang, Q.-R. Yuan, \& X. Zhou, et al.}
   \titlerunning{Multicolor Photometry of the Galaxy Cluster A98}
   \maketitle


%
%
\section{Introduction}
\label{sect:intro}



Following the hierarchical scenario of structure formation,
massive clusters form through episodic mergers of subunits, such
as groups and poor clusters, and through the continuous accretion
of field galaxies along the filaments (Zeldovich \etal 1982; West
\etal 1991, 1995; Colberg \etal 2000). Both the X-ray and optical
surveys have revealed a significant level of substructure in rich
galaxy clusters (Rhee et al. 1991; Forman \& Jones 1982; Beers et
al. 1991; Sarazin et al. 1992; Henry \& Briel 1993; Burns et al.
1994). Numerical simulation of the evolution of galaxy clusters
indicates that at least $50\%$ of apparently relaxed clusters
contain significant substructures (Salvador-Sole \etal 1993). The
dynamics of these ``lumpy'' clusters thus provides a means for
exploring cluster evolution, which may shed light on the theories
of large-scale structure formation (Kauffmann et al. 1999).

A98 is a good example of clusters with multiple components. It is
a rich (R=3), quite distant ($z=0.1042$, D=5) cluster (Abell et
al. 1989) without the large cD galaxy at its center (the
Bautz-Morgan type -BM:II-III). It was first selected optically by
Abell (1958), and was regarded as a single galaxy cluster before
the launch of the {\it Einstein} Observatory (Duus \& Newell
1977). The X-ray surface brightness distribution of A98 exhibits
two large enhancements which were considered to be associated with
the north and south components (Forman et al. 1981;  Henry et al.
1981; Jones \& Forman 1999). From then, A98 was extensively
studied as a typical double cluster. Dressler (1978a; 1978b)
analyzed the galaxy distribution of A98, and suggested that it
consists of two subclusters, namely the south (A98S) and the north
(A98N) ones. The radial velocities for galaxies in A98 were
discussed by many authors (Faber \& Dressler 1977; Dressler 1978b;
Beers et al. 1982; Zabludoff et al. 1990). However, the number of
spectroscopically confirmed member galaxies in A98 in previous
studies is very limited. Based on only 24 member galaxies, Beers,
Geller, \& Huchra (1982, hereafter BGH) estimated the virial
masses, mass to luminosity ratios of A98N and A98S. By using the
two-body model, they further calculated the probability of
gravitational binding, and derived that the two subclusters would
merge in another 3 billion years. Krempe\'{c}-Krygier \& Krygier
(1995, hereafter KK95) studied the dynamics of A98 on the basis of
only 29 cluster members. As a rich cluster containing wide-angle
tailed (WAT) radio galaxies, A98 is included in the sample of
WAT-containing clusters in Pinkney et al. (2000, hereafter PBLGH).
The redshifts of some galaxies in the A98 region were obtained in
that paper, but they did not investigate the dynamics of A98 in
detail.

Substructures in the optical surface density and radial velocity
are the typical signatures that allow to identify merging
clusters.  In this paper, we collect 74 cluster galaxies with
known spectroscopic redshifts from literature, and investigate the
dynamical substructures in A98. For a better understanding of
dynamics of A98, the faint member galaxies ($18.0<m_V<20.0$)
should be taken into account. This paper will present a multicolor
photometry of the galaxies in A98 with the
Beijing-Arizona-Taiwan-Connecticut (BATC) system. Based on the
spectral energy distributions (SEDs) of faint galaxies, we try to
supplement a large number of new member galaxies by applying the
photometric redshift technique. The enlarged data set may verify
the spatial distribution and dynamical properties of A98 to an
unprecedented depth. Additionally, the star formation histories of
the bright member galaxies may help us to understand the evolution
of A98.

This paper is organized as follows: In section2, we present the
BATC multicolor photometric observations and data reduction. In
section 3, we analyze the galaxies with known spectroscopic
redshifts in the A98 field, and distinguish a new galaxy group at
$z=0.12$ from the main concentration of A98. Section 4 presents
the SED selection of faint member galaxies of A98. In section 5,
the dynamical substructures and star formation properties are
investigated on the basis of the sample of spectroscopically
confirmed members and the enlarged sample of member galaxies.
Finally, we summarize our work in section 6. Throughout this paper
we assume the cosmological parameters
$H_0=70$kms$^{-1}$Mpc$^{-1}$, $\Omega_m=0.3$ and
$\Omega_\Lambda=0.7$.

\section{Observation and Data Reduction}
\label{sect:Obs}

The BATC multicolor photometric survey is designed to obtain the
optical SED information of the faint objects without spectroscopic
observation using the 60/90 cm f/3 Schmidt telescope of the
National Astronomical Observatories, Chinese Academy of Sciences
(NAOC), located at Xinglong site with an altitude of 900m. Before
October 2006, an old Ford CCD camera with a format of
2048$\times$2048 was mounted at the main focus of the telescope.
The field of view was 58'$\times$58', with a scale of 1."7
pixel$^{-1}$.  For pursuing a better spatial resolution and a
higher sensitivity in blue bands, a new E2V 4096 $\times$ 4096 CCD
camera was equipped. The field of view becomes larger (92'
$\times$ 92') with a spatial scale of 1."35 pixel$^{-1}$. The
pixel sizes for the old and new CCD cameras are 15${\mu}m$ and
12${\mu}m$, respectively, and the pixel size ratio is exactly 5:4.
The newly equipped CCD camera has a high quantum efficiency of
92.2\% at 4000 \AA. The BATC filter system contains 15
intermediate-band filters covering the wavelength range from 3000
to 10000\AA. These filters were especially designed to avoid
bright night sky emission lines (Fan et al. 1996). The
transmission curves can be found in Yuan et al. (2003) and Xia et
al.(2002).

From 1996 to 2006,  we accumulated 37 hours in only 12 bands, from
$d$ to $p$, with the old CCD camera. In recent two years the
exposures in a,b,c filters were completed with the new CCD camera.
The total exposure time reaches more than 42 hours (see the
observational statistics in Table 1). With an automatic
data-processing software, PIPELINE I (Fan et al. 1996), we carried
out the standard procedures of bias subtraction, flat-field
correction, and position calibration. The technique of integral
pixel shifting is used in the image combination during which the
cosmic rays and bad pixels were removed by comparing multiple
images.

For detecting and measuring the flux of sources within a given
aperture in the BATC images, we use a photometry package, PIPELINE
II, developed on the basis of DAOPHOT kernel (Zhou et al. 2003a),
to perform aperture photometry. An object is considered to be
detected if its signal to noise ratio is larger than the threshold
3.5 $\sigma$ in $i$, $j$, and $k$ bands. Thanks to the pixel size
ratio between the old and new CCDs is 5:4, we adopt a radius of 4
pixels as the photometric aperture for the images in 12 bands
(from $d$ to $p$), and a radius of 5 pixels for the images in
other three bands (from $a$ to $c$). The flux calibration in the
$h$ band is performed using the Oke-Gunn primary flux standard
star HD 19445, HD 84937, BD+26 2606, and BD+17 4708 (Gunn \&
Stryker 1983). To achieve the {\it relative} SEDs of the sources
detected by the BATC system, Zhou et al. (1999) developed a method
of model calibration on the basis of the stellar SED library. No
calibration images of the standard stars are needed during the
flux calibration. Using this model calibration method, as a
result, the SEDs of about 9000 sources have been obtained for
further analysis.

The magnitudes within a fixed photometric aperture is somewhat
different from the total magnitudes of galaxies given in some
catalogs. For assessing the measurement errors at specified
magnitudes, we separate stars into different bins of magnitudes
with an interval of 0.5 mag, and we find that magnitude error in
each filter are larger at fainter depths. A typical error is less
than 0.02 mag for the stars brighter than 16.5 mag, and about 0.05
mag for the stars with V $\sim$ 18.5 mag.


\input{tab1.tex}

\section{Discovery of a new galaxy group 'A98X'}

\subsection{Distribution of spectroscopic redshifts }
For studying the dynamics of the galaxy cluster A98, 122 galaxies
with known spectroscopic redshifts ($z_{sp}$) in the $58' \times
58'$ field centered at A98 are extracted from NASA/IPAC
Extragalactic Database (NED). Most of the spectroscopic data were
contributed by Struble \& Rood (1999) who present a compilation of
the redshifts for 1572 rich clusters in Abell, Corwin \& Olowin
(1989). Fig. 1a shows the distribution of spectroscopic redshifts
of these bright galaxies. The highest peak, centered at $z_{sp}
\sim 0.104$, is isolated and less contaminated. So it is
unambiguous to regard the 74 galaxies with $0.095 < z_{sp} <
0.115$ as the spectroscopically confirmed member galaxies in A98,
and we refer to these member galaxies as `sample I'. Table 2 lists
the information of the position and spectroscopic redshift of
these 74 galaxies.

\input{tab2.tex}

\begin{figure}[t]
\centering
\includegraphics[width=135mm]{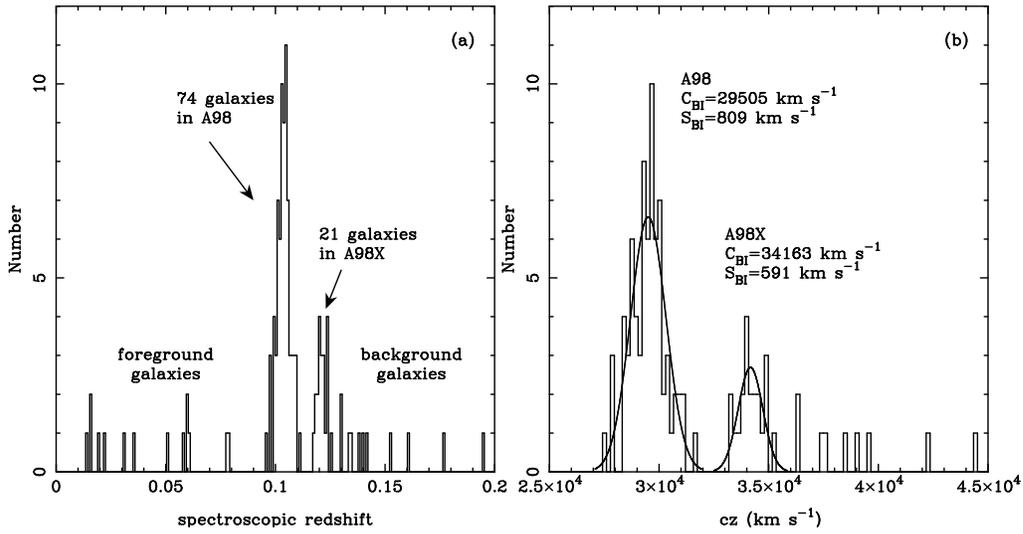}
\caption{Left: Distribution of spectroscopic redshifts for 122 known
galaxies in the A98 region. The bin size is 0.00125. Right: Distribution of radial velocities for the galaxies in A98 and A98X, with a Gaussian fitting superposed on each component.}
\label{fig1}
\end{figure}


To quantify the velocity distribution of member galaxies, we
firstly convert the redshifts into true cosmological velocities by
$ v=c[(1+z_{sp})^2-1]/[(1+z_{sp})^2+1] $ , where $c$ is the light
speed. Then, we use the ROSTAT software (Beers, Flynn, \& Gebhardt
1990) to calculate two resistant and robust estimators analogous
to the velocity mean and standard deviation, namely, the biweight
location ($C_{BI}$) and scale ($S_{BI}$). For these 74 member
galaxies, we achieve $C_{BI}= 29505\pm 94$ \kms and $S_{BI}=
809\pm 77$ \kms. Fig. 1b shows the distribution of the
line-of-sight velocities for these 74 galaxies. Taking a
cosmological correction factor of $(1+z)^{-1}$ into account, the
velocity dispersion of A98 should be $732 \pm 70$ \kms.

Our sample of member galaxies is much larger than the ones in BGH
and KK95. Based on only 29 member galaxies, KK95 derived a mean
velocity of 29757 \kms and a velocity dispersion of 859 \kms.
Using 69 member galaxies in A98, PBLGH derived the mean velocity
of $31085\pm96$ \kms and the velocity dispersion of $797 \pm 73$
\kms. Their values are larger than our estimates due to the
contamination of the galaxy group at $z=0.120$, which will be
discussed in the next subsection.

\subsection{The south clump `A98X' at $z=0.120$}
It is interesting to find in Fig. 1 that there are 21 galaxies
with the redshifts between 0.115 and 0.128. For this velocity
concentration, we obtain the biweight location of $C_{BI}=34163\pm
133$ \kms and the biweight scale of $S_{BI}= 591\pm 76 $ \kms.
Considering the cosmological correction, the velocity dispersion
of this peak is $528 \pm 68$ \kms. The separation between these
two velocity peaks is about $4658\pm 163$ \kms.

The left panel of Fig. 2 shows the spatial distribution for 74
member galaxies(denoted by asterisks) and 21 galaxies with the
redshifts between 0.115 and 0.128 (denoted by open triangles),
with respect to the NED-given central position of A98
($00^h46^m26.^s6$, $+20^{\circ}$29'23"; J2000.0). We superpose the
contour maps of surface density that has been smoothed by a
Gaussian window of $\sigma=1.'6$. The contour map shows that these
21 galaxies with $z_{sp} \sim 0.120$ belong to a separate clump,
about 15' south to the main concentration. As the separation
between the two peaks is very remarkable, we refer to the south
clump as 'A98X'.

To show the prominence of the clump A98X in both the velocity
space and the projected map, we make use of the $\kappa$-test
(Colless \& Dunn 1996) for the A98/A98X system as a whole. The
statistic $\kappa_n$ is defined to quantify the local deviation on
the scale of groups of $n$ nearest neighbors. A larger $\kappa_n$
indicates a greater probability that the local velocity
distribution differs from the overall velocity distribution. The
probability $P(\kappa_n>\kappa_n^{obs})$ can be calculated by
Monto Carlo simulations with random shuffling velocities. When the
scale of the nearest neighbors $n$ varies from 3 to 10, the
probability $P(\kappa_n>\kappa_n^{obs})$ is nearly zero, which
means the substructure appears very obvious at different scales.
The bubble plot at $n=6$ is given in the right panel of Fig. 2.
Since the bubble size is proportional to
$-\log[P_{KS}(D>D_{obs})]$, the clustering of larger bubbles at
(1.'55,-14.'43) can trace the significant substructure, which
corresponds to the galaxy group `A98X' centered at
($0^h46^m33.^s2$, $20^{\circ}14'57"$; J2000.0).

\begin{figure}[t]
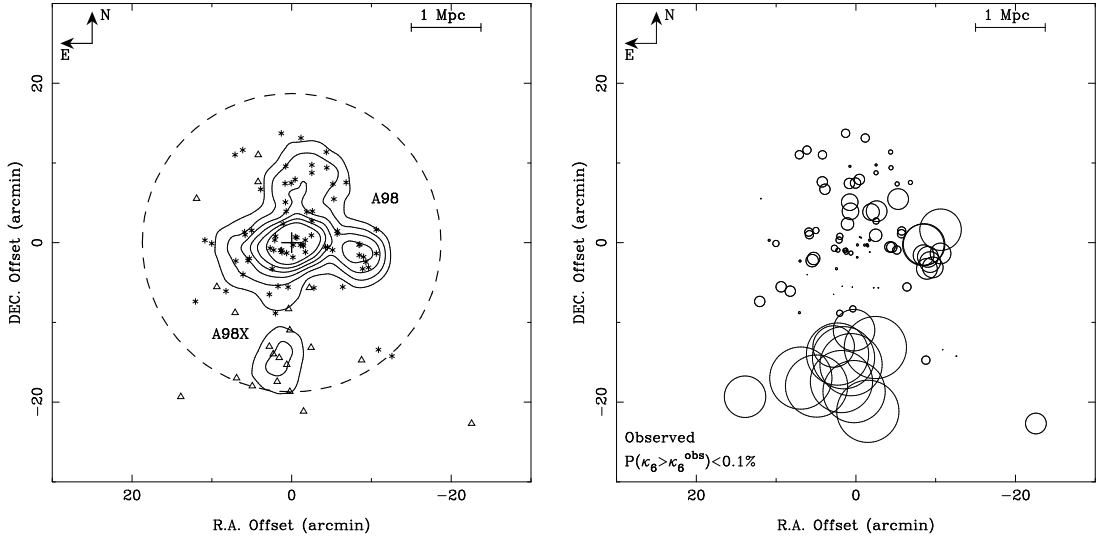

\includegraphics[width=70mm]{fig02a.eps}
\includegraphics[width=70mm]{fig02b.eps}
\begin{minipage}[]{145mm}
\caption{Left: Spatial distribution for 95 galaxies in A98
(denoted by asterisks) and A98X (denoted by triangles), superposed
by the contour map of surface density where the smoothing Gaussian
window with $\sigma = 1.'6$ is used. The contour levels are
0.09,0.15,0.21,0.27,0.33,0.39 $arcmin^{-2}$, respectively. The
dashed circle means a typical region of rich clusters with the
Abell radius of 1.5$h^{-1}$ Mpc; Right: Bubble plot for the 95
galaxies in the A98 and A98X components. }
\end{minipage}
\label{fig2}
\end{figure}

We fail to find any clusters and/or groups of galaxies from
literature and existing catalogs of galaxy clusters around the
position of the clump A98X. This concentration has not been
mentioned in any previous investigations. An interesting question
is whether the clump A98X is just a subcluster of A98 or a
newly-detected cluster/group of galaxies. To answer this question,
it is necessary to verify whether A98X is gravitationally bound to
the A98 cluster or not.

\subsection{The Application of Gravitational Binding Criterion}

The masses of the main concentration of A98 and the clump A98X can
be estimated by applying virial theorem. Assuming that each
cluster is bound and the galaxy orbits are random, the virial mass
$(M_{vt})$ can be derived from the following standard formula
(Geller \& Peebles 1973; Oegerle \& Hill 1994):
\begin{equation}
M_{vt}=\frac{3\pi}{G}\sigma_{r}^{2}DN_{p}
\left(\sum\limits_{i>j}^{N}\frac{1}{\theta_{ij}}\right)^{-1},
\end{equation}
where $\sigma_{r}$ is the line-of-sight velocity dispersion, D is
the cosmological distance of the cluster, $N_{p}=N(N-1)/2$ is the
number of galaxy pairs, and $\theta_{ij}$ is the angular
separation between the galaxies $i$ and $j$. The virial masses of
$9.12 \times 10^{14}M_{\odot}$ and $7.98 \times 10^{14}M_{\odot}$
are derived for A98 and A98X, respectively. Then we specify the
limits of the bound solutions by using Newtonian criterion of
gravitational binding (Beers et al. 1982):
\begin{equation}
V_{r}^{2}R_{p}\leq2GM\sin^{2}\alpha\cos\alpha,
\end{equation}
where $V_{r}$ is the relative velocity along the line of sight,
$R_{p}$ is the projected separation, $M$ is the total mass of the
two clusters, and $\alpha$ is the angle between the plane of the
sky and the line connecting the two clusters. The projected
separation of A98 and A98X is $R_{p}=1.78$ Mpc, and the actual
observed $V_{r} \sim 4200 \pm 163 $km$s^{-1}$ in the rest frame of
A98. The resulting constraints on $V_{r}$ for bound orbits are
shown in Fig. 3, as a function of the projection angle, $\alpha$.
The solid curve in this plot separates the regions of bound and
unbound. As the figure shows, the maximum $V_{r}$ value for the
bound solution is $\sim1800$ \kms, which is obviously much lower
than the observed $V_{r}\sim 4200 \pm 163 $ \kms. There is no
doubt that the clump A98X containing 21 galaxies is
gravitationally unbound to the cluster A98. Our conclusion is that
the A98X is a separate group of galaxies. In our following
dynamical analysis of A98, this galaxy group will be excluded.

It should be noted that the main concentration of A98 is not a
well virialized system which contains more than two subclumps (see
section 5). It might be arbitrary to derive the mass of main
concentration via the virial theorem. However, we believe that the
probable bias in our mass estimate is not significant enough to
influence above conclusion.

\begin{figure}[t]
\centering
\includegraphics[width=75mm]{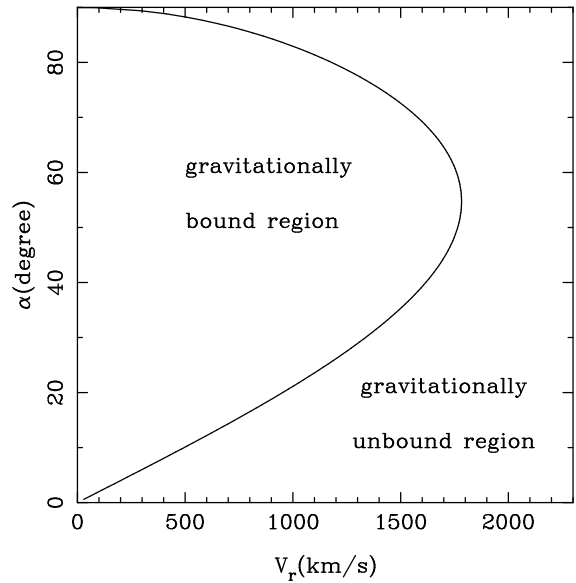}
\begin{minipage}[]{145mm}
\caption{Projection angle $\alpha$ as a function of radial
velocity $V_{r}$ given by the total mass
$1.71\times10^{15}M_{\odot}$. The solid curve separates the
bound and unbound regions. }
\end{minipage}
\label{fig3}
\end{figure}

\section{SED Selection of Faint Cluster Galaxies}

\subsection{Star-Galaxy Separation}
For the galaxies with known spectroscopic redshifts given by the
NED, we cross-identified with the BATC-detected sources. All the
sources in the BATC catalog within the searching circle (defined
by a radius of 5'') centered on the NED-given galaxies were
extracted. The identification is rather unambiguous. For the case
of several counterparts within the searching area, we pick up the
brighter BATC source as the right counterpart. As a result, 118
bright galaxies with known $z_{sp}$ values are identified. The
15-band SEDs of these galaxies will be used to check the
reliability of our photometric redshift technique.

For selecting the probable faint member galaxies from the
remaining BATC sources, we firstly perform the star/galaxy
separation. As shown in Yuan et al.(2001), the color-color diagram
is a powerful tool of classification. Since the spectra of
redshifted galaxies differ significantly from those of stars,
different regions in the color-color diagram are occupied by
different classes of objects. Fig. 4 shows two color-color
diagrams used for our star-galaxy separation. The diagrams include
the following categories of sources: (1) all types of stars in our
SED template library (denoted by filled triangles), (2)
morphologically various galaxies with template SEDs (denoted by
open circles), (3) the spectroscopically confirmed member galaxies
of A98 (denoted by crosses), and (4) all the remaining sources
detected by the BATC photometry (denoted by dots). The filters
$a$(3360\AA ),$b$(3890\AA ),$h$(6075\AA ), and $p$(9745\AA ) are
used in the diagrams. It can be seen in Fig. 4 that the stars in
the SED template library lie in a well-defined zone stretching
from top left to bottom right, while the confirmed galaxies
distribute just above the zone. Taking the dashed lines in Fig.4
as the boundaries of star-galaxy separation, we pick out the
galaxies simultaneously detected in at least 11 BATC bands. As a
result, we obtained 1490 faint galaxies that are located within
the boundries in both panels, to which we shall apply the
photometric redshift technique.

\begin{figure}[t]
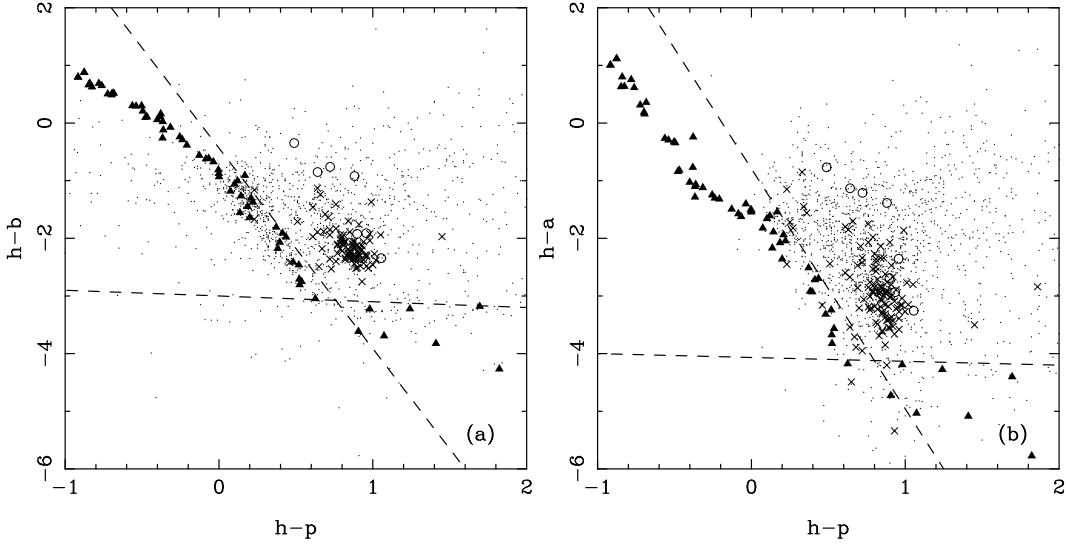

\includegraphics[width=70mm]{fig04a.eps}
\includegraphics[width=70mm]{fig04b.eps}
\caption{Color-color diagrams used for star-galaxy separation (
Asterisks for all types of stars in the SED template library, open
circles for various galaxies with template SEDs, crosses for the
         confirmed galaxy, small dots for the detected sources). The
         dashed lines are taken as boundaries of separation.}
\label{fig4}
\end{figure}

\subsection{Photometric Redshifts and Cluster Membership}
The technique of photometric redshift can be used to estimate the
redshifts of galaxies by using the SED information covered a wide
range of wavelength instead of the spectroscopy. This technique is
extensively applied to the multicolor photometric surveys for
detecting the faint and distant galaxies (Pell\'{o} et al. 1999a,
b; Bolzonella et al. 2000; Rowan-Robinson et al. 2008; Ilbert et
al. 2009) and for selecting the cluster galaxies (Brunner \& Lubin
2000; Finoguenov et al. 2007). For a given object, based on the
standard SED-fitting code called HYPERZ (Bolzonella, Miralles, \&
Pell\'{o}¡ä 2000), the photometric redshift, $z_{ph}$, corresponds
to the best fit (in the $\chi^{2}$-sense) between its photometric
SED and the template SED generated by convolving the galaxy
spectra in template library with the transmission curves of BATC
filters. Previous work has evaluated the accuracy of photometric
redshift with the BATC multi-band data (Yuan et al. 2001, 2003;
Zhou et al. 2003b; Yang et al. 2004).  In our SED fitting, only
the normal galaxies are taken into account in the reference
templates. The dust extinction with a reddening law of the Milky
Way (Allen 1976) is adopted, and $A_V$ is allowed to be flexible
in a range from 0.0 to 0.2, with a step of 0.02. The photometric
redshift of a galaxy with the BATC SED is searched from 0.0 to
0.5, with a step of 0.005.

For the 118 galaxies with known spectroscopic redshifts, a
comparison between the photometric redshifts $z_{ph}$ and the
spectroscopic redshifts $z_{sp}$ is shown in Fig. 5. The solid
line corresponds to $z_{ph}=z_{sp}$, and the dashed lines indicate
an average redshift deviation of 0.023, and the error bar of
$z_{ph}$ corresponds to $68\%$ confidence level in photometric
redshift determination. It is obvious that our $z_{ph}$ estimate
is basically consistent with the spectroscopic redshift. For the
74 member galaxies in sample I, the biweight location ($C_{BI}$)
and scale ($S_{BI}$) of the $z_{ph}$ estimate are 0.097 and 0.008,
respectively. There exists a slight systematic offset in the
$z_{ph}$ distribution, with respective to the $z_{sp}$
distribution. Taking the selection criterium of $2\sigma$
clipping, 65 member galaxies (about 90 percent) are found to have
their photometric redshifts in a range from 0.081 ($=0.097-2
\times 0.008$) to 0.113 ($=0.097 + 2\times 0.008)$. This $z_{ph}$
region can be applied as a selection criterium in the following
membership determination for the faint galaxies detected by the
BATC multicolor photometry only.

\begin{figure}[t]
\includegraphics[width=70mm]{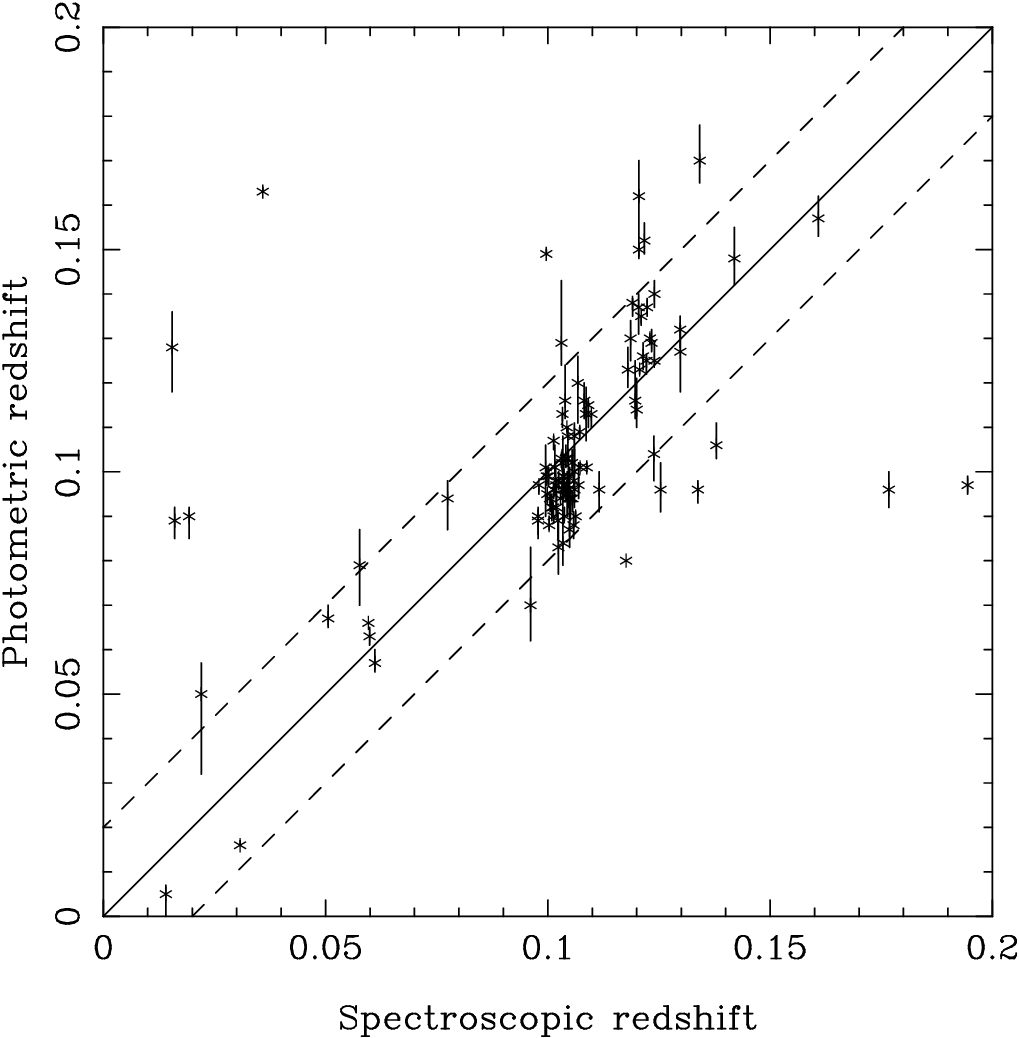}
\includegraphics[width=70mm]{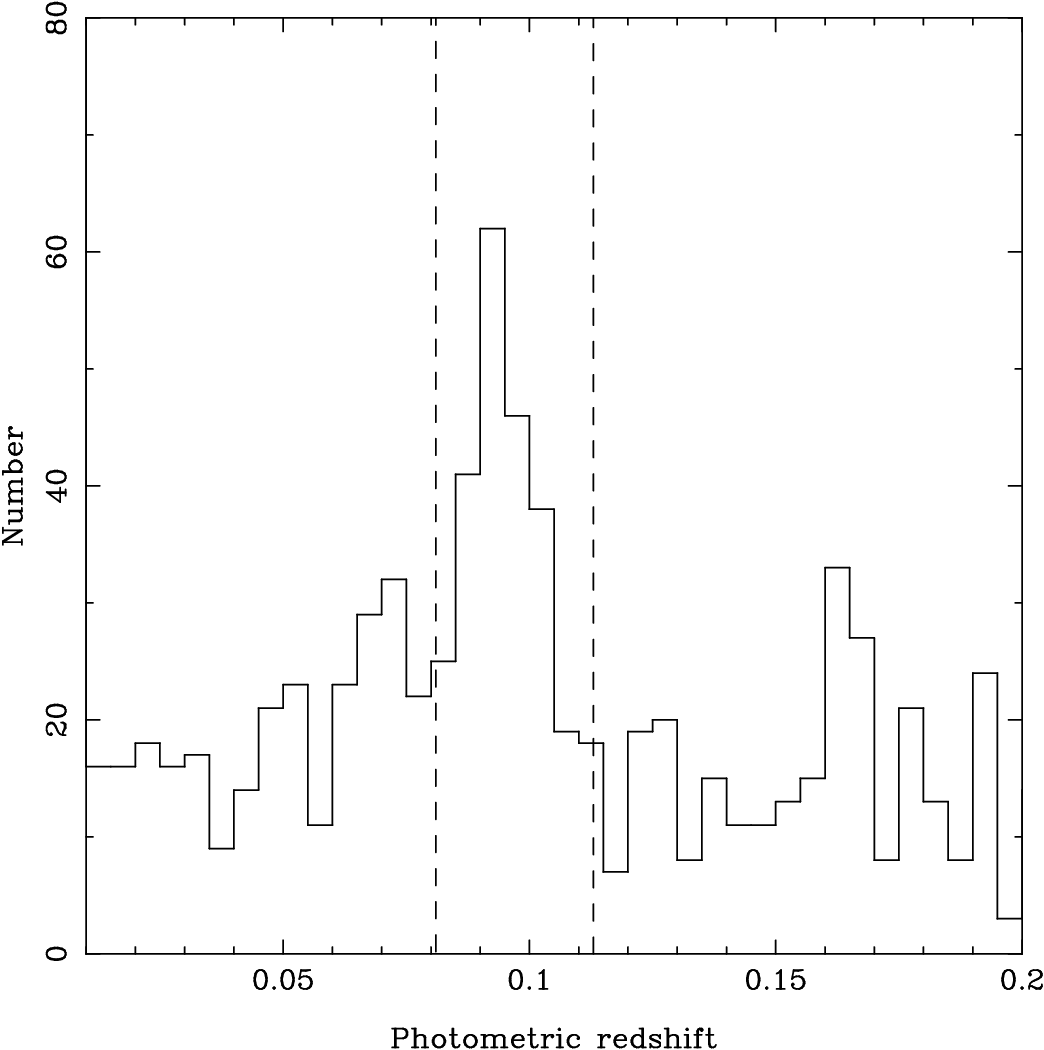}
\begin{minipage}[]{74mm}
\caption{Comparison between photometric redshift $z_{ph}$ and
        spectroscopic redshift $z_{sp}$ for 118 galaxies in
        the field of view. The solid line corresponds to $z_{ph}=
        z_{sp}$, and the dashed lines indicate an average deviation
        of 0.023.}
\end{minipage}\label{fig5}
\begin{minipage}[]{74mm}
\caption{Distribution of photometric redshifts for the galaxies
($z<0.2$) . The dashed lines are plotted as the photometric
redshift range of cluster member candidates.}
\end{minipage}
\label{fig6}
\end{figure}


Fig. 6 shows the histogram of photometric redshifts for faint
galaxies with $z_{ph}<0.2$ in the viewing field. The galaxies with
$0.081 < z_{ph} < 0.113$ within one Abell radius (1.5$h^{-1}$Mpc),
corresponding to 18.71 arcmin at $z=0.104$, are selected as member
candidates of A98. As a result, there are 198 faint member
candidates, among which 137 galaxies are regarded as early-type
galaxies and 61 galaxies are regarded as late-type galaxies by the
SED-fitting procedure.

It is well known that there exists a correlation between the color
and absolute magnitude for the early-type galaxies (C-M relation,
see Bower et al. 1992), in the sense that the bright galaxies are
redder, which can be used for verifying the membership selection
of the early-type galaxies. For selecting the early-type galaxies,
We take the morphological types of the best-fit SED templates.
Fig. 7 presents the correlation between the color index $b-h$ and
magnitude in $h$ bandpass for 203 early-type member galaxies,
including 66 early-type galaxies with known $z_{sp}$ and 137
newly-selected early-type galaxies. The solid line denotes the
linear fitting of the 66 spectroscopically confirmed member
galaxies: $b-h=-0.10(\pm0.04)h +3.89(\pm 0.79)$, and the dashed
lines represent 1 $\sigma$ deviation. As shown in Fig. 7, most
majority of the early-type candidates agree with the C-M relation
derived by 66 bright early-type galaxies. There is no faint
early-type candidates with the color $b-h$ beyond the 2 $\sigma$
deviation of intercept. By combining the 198 newly-selected
members and 74 spectroscopically confirmed member galaxies in
sample I, we obtain an enlarged sample of 272 member galaxies,
which is referred to as sample II.

Table 3 presents the catalog of SED information for the 198 newly
selected members, as well as the celestial coordinate, photometric
redshift, and morphological class of the best-fit template. The
classification indices, $T$, ranging from 1 to 7, are defined to
denote E, S0, Sa, Sb, Sc, Sd, and Im galaxies, respectively.

\begin{figure}[bt]
\centering
\includegraphics[width=70mm]{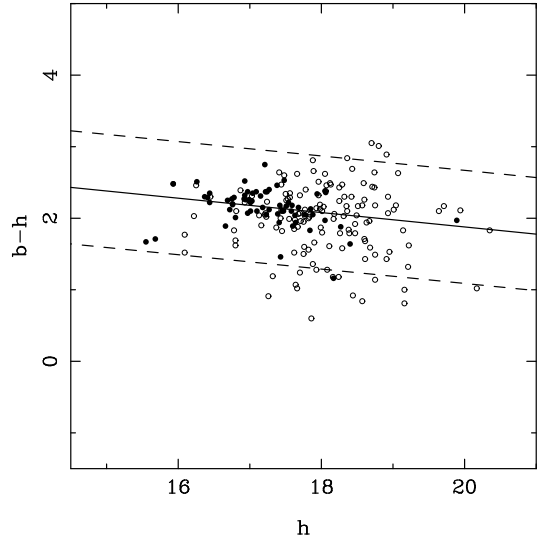}
\caption{Color-magnitude correlation for 203 early-type galaxies
in A98. Filled circles denote 66 early-type member galaxies which have
been spectroscopically confirmed, and open circles denote
137 newly-selected member galaxies. The solid line shows the
linear fit for 66 bright galaxies, and the dashed lines
correspond to 1 $\sigma$ deviation.}
\label{fig7}
\end{figure}

\section{The Properties of The Cluster A98}

\subsection{Projected Distribution of Cluster Galaxies}
Fig. 8 shows the projected position of the galaxies in samples I
and II, superposed with the contour maps of surface density where
the smoothing Gaussian window of $\sigma= 1.'6$ is used. 74 member
galaxies with known spectroscopic redshifts are denoted by filled
circles and 198 photometrically selected galaxies are denoted by
open circles. Two X-ray peaks given in Forman et al.(1981) are
also marked with filled triangles. Fig.8a shows that the 74 bright
member galaxies of A98 seem to deviate from spherically symmetric
distribution. The contour map of surface density appears elongated
in the north-south direction. Two X-ray peaks are found to be
associated with the surface density substructures A98N and A98S.
Previous spectroscopy shows that the subcluster A98N is much
poorer than the main concentration A98S, which is remarkable in
Fig. 8a. Our BATC multicolor photometry facilitates the finding of
large number of faint member galaxies around the northern
subcluster A98N, which makes the northern substructure in surface
density more remarkable (see Fig. 8b).

What we should keep in mind is that the surface density of A98N is
still lower than that of A98S, even after we take all the faint
member galaxies into account. However, the northern subcluster
A98N is brighter than the southern A98S in the X-ray emission.
Forman et al. (1981) reported that the flux ratio between A98N and
A98S is about 3:2. Henry et al. (1981) derived a ratio of X-ray
luminosity between A98N and A98S of 1.3:1. Considering the
contribution in X-ray surface brightness of the central galaxy
0043+2020, KK95 fitted with the $\beta$ model, and obtained a
lower X-ray luminosity ratio of 1.1:1.


Apart from the northern subcluster A98N, a western subcluster can
be seen in Fig. 8, at about 12 arcmin west of A98S. This
subcluster has not been mentioned before, and we refer to it as
A98W. After adding 198 photometrically selected galaxies, Fig. 8b
shows that A98N appears more significant and the western clump
A98W still exists. In general, the projected position distribution
of the galaxies in sample II is consistent with that of the bright
galaxies in sample I.

\begin{figure}[htb]
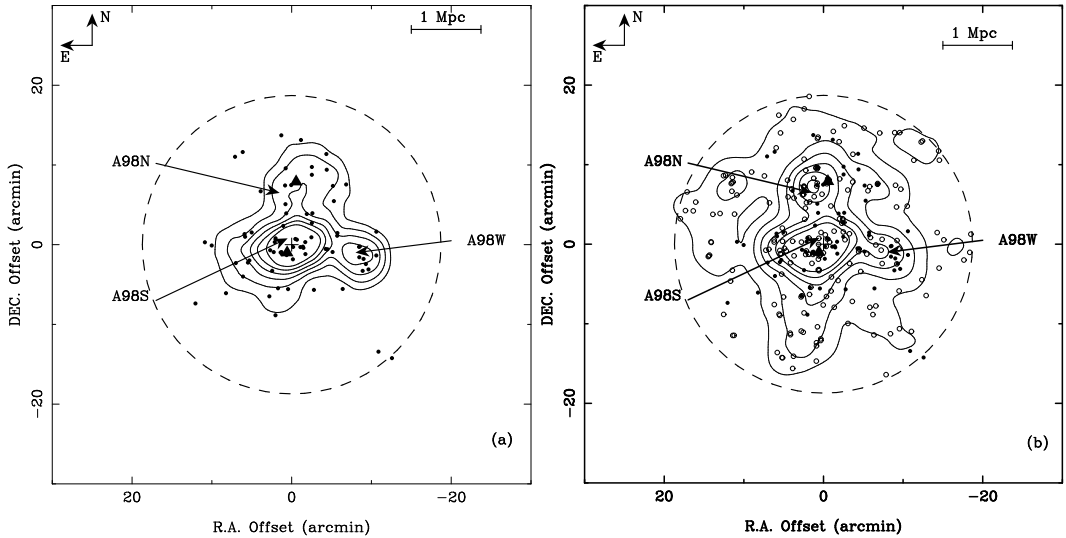

\includegraphics[width=70mm]{fig08a.eps}
\includegraphics[width=70mm]{fig08b.eps}
\caption{Left (a): The positions of 74 spectroscopic member
galaxies of A98 in sample I. The surface density is smoothed with
a Gaussian window of $\sigma=1.'6$.  Contour map is superposed
with the surface density levels 0.09,0.15,0.21,0.27,0.33,0.39
arcmin$^{-2}$. The dashed circle means a typical region of rich
clusters with the Abell radius of 1.5$h^{-1}$ Mpc. Two X-ray peaks
are marked with filled triangles. Right (b): Position distribution
of the  272 member galaxies in sample II, including 74
spectroscopically confirmed member galaxies (filled circles) and
198 newly-selected galaxies (open circles). The contour map of the
surface density for these galaxies are also given with contour
levels 0.13,0.28,0.43,0.58,0.73,0.88 arcmin$^{-2}$.}
\label{fig8}
\end{figure}

\subsection{Localized Velocity Structure}
However, the clumps mentioned above might be the enhancement
simply due to projection effect. If a cluster merger occurs along
the direction with a definite projection angle, say $\alpha >
20^{\circ}$, with respective to the plane of sky, the
substructures can be detected by mapping the localized variation
in velocity distribution (Colless \& Dunn 1996). The $\kappa$-test
defines a test statistic ${\kappa}_n$ to characterize the local
deviation on the scale of groups of $n$ nearest neighbors.


For detecting the substructures in A98, we perform the
$\kappa$-test for the galaxies in samples I and II. For sample I,
there is no significant substructure detected in the localized
velocity. We performed $10^{3}$ simulations to estimate
probability $P$($\kappa_{n}> \kappa^{obs}_{n}$) in all cases. The
probability $P$($\kappa_{n}> \kappa^{obs}_{n}$) is found to be
more than 5\% in a wide range of neighbor sizes, which means that
no substructure is detected at 2$\sigma$ significance. Table 4
gives the result of our $\kappa$-test for the galaxies in sample
I. However, for the enlarged sample of 272 member galaxies, the
probability of substructure detection is greater. Considering the
larger relative error in photometric redshift determination, the
probability estimate of sample II might be untrue.

\setcounter{table}{3}
\begin{table}[htbp]
\bc
\caption[]{Result of $\kappa$-Test for 74 spectroscopically
confirmed member galaxies}
\begin{tabular}{ccccccccc}   \hline
\noalign{\smallskip}
Neighbor size $n$ & 3& 4& 5& 6& 7& 8& 9 \\
\hline \noalign{\smallskip}
$P(\kappa_n>\kappa_n^{\rm obs})
$ &11.4.2\%& 5.3\%& 9.2\%& 14.4\%& 16.6\%& 24.9\% & 40.6\%\\
\noalign{\smallskip}   \hline
\end{tabular} \ec
\end{table}

Bubble plots in Fig. 9 show the localized velocity variation for
sample I and sample II, using 6 nearest neighbors. The bubble size
for each galaxy is proportional to $-log[P_{KS}(D>D_{obs})]$. For
the galaxies in sample I (see Fig. 9a), there are indeed bunches
of bubbles at the positions of subcluster A98S and A98W, though
the bubble sizes are not very large. A close comparison between
Fig. 8a and Fig. 9a indicates that A98S and A98W are not simply
due to projection effect, they are most likely to be real
substructures. On the other hand, no bubble clustering is found
for the northern subcluster A98N. It seems to support the two-body
model that we are looking at a cluster merger between A98N and
A98S occurred largely in the plane of the sky, which is consistent
with the small projection angle derived in KK95.

For the galaxies in sample II (see Fig. 9b), the clustering of
bubbles at A98N is enhanced. Taking the faint member galaxies into
account, the localized velocity variation between A98S and A98N
appears more remarkable. Within the central regions of A98N and
A98S, defined by the contour curve at 0.58 arcmin$^{-2}$ in Fig.
8b, there are 28 and 73 galaxies, respectively. For the 28
galaxies in A98N, the biweight location is $27029\pm 335$ \kms,
However, for the 73 galaxies in A98S, we obtain a higher biweight
location, $28721 \pm 207$ \kms. The difference in radial velocity
between A98N and A98S is about $1692 \pm 394$ \kms, which is
larger than the typical velocity dispersion of a rich cluster.
Fig. 10 gives the distributions of radial velocities for these
galaxies. The large velocity difference seems to be a challenge
for the bound system. It is noteworthy that the photometric
redshifts for the faint member galaxies have a lower precision,
and it is necessary to verify the inertal dynamics between A98N
and A98S by the follow-up deep spectroscopic observation. There
are only 6 and 27 galaxies with known $z_{sp}$ values within the
central regions of A98N and A98S, respectively, for which the
velocity distributions are shown in the shadowed histograms in
Fig. 10. For the spectroscopic subsamples in A98N and A98S central
regions, the biweight locations are 28943 and 29609 \kms,
respectively. The velocity difference becomes about 666 \kms, much
smaller than the value of enlarged samples, which supports the
gravitationally bound nature of these two subclusters.

\begin{figure}[tb]
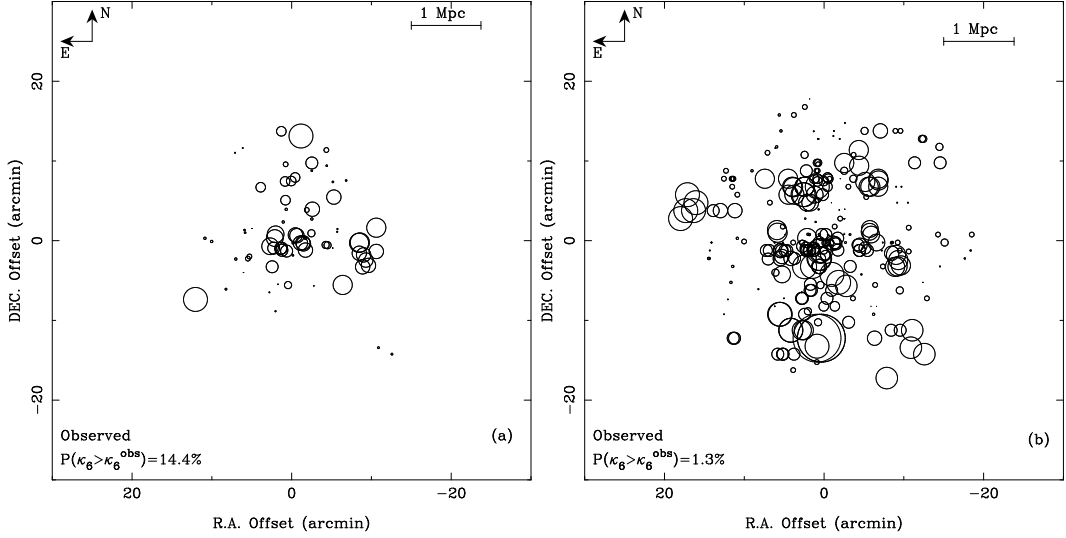

\includegraphics[width=70mm]{fig09a.eps}
\includegraphics[width=70mm]{fig09b.eps}
\caption{(a)Bubble plot for 74 spectroscopic member galaxies of
A98. It shows the localized variation for groups of the six
nearest neighbors. (b)Bubble plot for all the 272 member galaxies
of A98. The same group size as Fig.(a) is used to characterize the
localized variation.}
\label{fig9}
\end{figure}

\begin{figure}[tb]
\centering
\includegraphics[width=70mm]{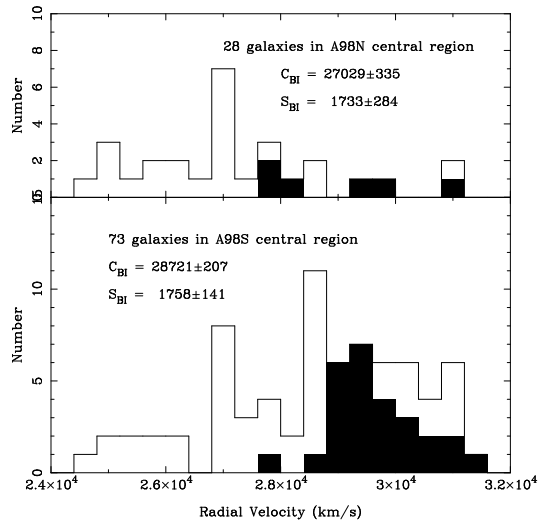}
\caption{Velocity distribution of the galaxies in the central
regions of A98N and A98S, defined by the contour curve at 0.58
arcmin$^{-2}$ in Fig. 8b. The velocity distributions of the
galaxies in spectroscopic subsamples are shown by the shadowed
histograms. }
\label{fig10}
\end{figure}

\subsection{Star Formation Properties}

Undoubtedly, the star formation histories of the member galaxies
may shed some light on the evolution of their host cluster. With
the evolutionary synthesis model, PEGASE (version 2.0, Fioc \&
Rocca-Volmerange 1997; 1999), we study the star formation
properties of A98. Assuming a Salpeter (1955) initial mass
function (IMF) and a star formation rate (SFR) in exponentially
decreasing form, $SFR(t) \propto e^{-t/{\tau}}$, where the time
scale $\tau$ ranges from 0.5 to 30.0 Gyr. In order to avoid the
degeneracy between age and metallicity in the model, we adopt the
same age of 12.2 Gyr, responding to the age of first generation
stars at $z=0.104$, for all member galaxies in A98. A zero initial
metallicity of interstellar medium (ISM) is taken. As a result, a
series of rest-frame modelled spectra with various star formation
histories are generated by running the PEGASE code, and then they
are redshifted to the observer frame for a given spectroscopic
redshift. Convolved with the transmission functions of all the
BATC filters, the template SED library for the BATC multicolor
photometric system (i.e., relative apparent magnitudes at 15 BATC
fileters) can be obtained.

Based on the template SED library, we search for the best fit (in
the $\chi^2$ sense) of the observed SEDs of 74 member galaxies
with known spectroscopic redshifts. The SFR time scale ($\tau$),
mean ISM metallicity ($Z_{ISM}$), and mean stellar age
($t_{\star}$) can be derived for each bright galaxy. Fig. 11
presents the star formation properties as a function of the local
surface density, $\Sigma$, which is defined by the number of
galaxies within an area with a radius of 2.5 arcmin. It is clear
in Fig. 11 that the star formation property of the member galaxies
in A98 are found to be dependent upon the local density. Panel (a)
shows that the galaxies in the outer region are likely to have a
longer SFR time scales than those in the core region. Considering
that the late-type galaxies tend to have longer time scales of
star formation, our result is consistent with the
morphology-density relation first pointed out by Dressler (1980),
which can be well explained in the context of hierarchical
cosmological scenario (Poggianti 2004). As shown in the panels (c)
and (d), the outlier member galaxies are likely to possess younger
stellar population, which results in a smaller mean stellar age
weighted by either mass or light.

Fig. 11b gives the variation of the mean ISM metallicities with
their local densities. The outlier galaxies have higher
probability to have a lower mean ISM metallicity. It is regarded
that the galaxies in the core region tend to be more massive and
luminous. The underlying physical correlation that can interpret
Fig. 11b is the luminosity-metallicity relation (Lequeux et al.
1979; Melbourne \& Salzer 2002) and the mass-metallicity relation
(e.g., Garnett \& Shields 1987; Tremonti et al. 2004). Fig. 12
gives the SFR time scales $\tau$ and the mean ISM metallicities as
the functions of the magnitude in $h$ band. As all galaxies in A98
have the same distance modulus, the apparent magnitudes could
reflect their intrinsic luminosities. For the bright and massive
cluster galaxies in the core region of A98, their star formation
activities have been reduced by some physical processes via
environmental effects, such as galaxy-galaxy interaction,
harassment, gas stripping, strangulation (Poggianti 2004; Yuan et
al. 2005), which leads to a short SFR time scale. From Fig. 11b
and Fig. 12b, no bright ($m_h<16.5$) and central ($\Sigma
> 0.4$ gal. arcmin$^{-2}$) galaxies are found to have the mean ISM
metallicity less than 0.035, which is consistent with the ideas
that more massive galaxies form fractionally more stars in a
Hubble time than the low-mass counterparts, and metals are
selectively lost from the faint galaxies with shallow potential
wells via galactic winds (Tremonti et al. 2004).

We also try to find any trends of the star formation property
along the distance from the main concentration of A98. However, no
significant environmental effect is found, indicating that A98 is
a dynamically complex cluster, and the clustercentric distance is
not a good environmental indicator. An alterative explanation is
that the star formation activities of the galaxies in a cluster
with ongoing merger events might be more sensitive to the
galaxy-scale gravitational interaction, not to the cluster-scale
environment.

\begin{figure}[h]
\centering
\includegraphics[width=120mm]{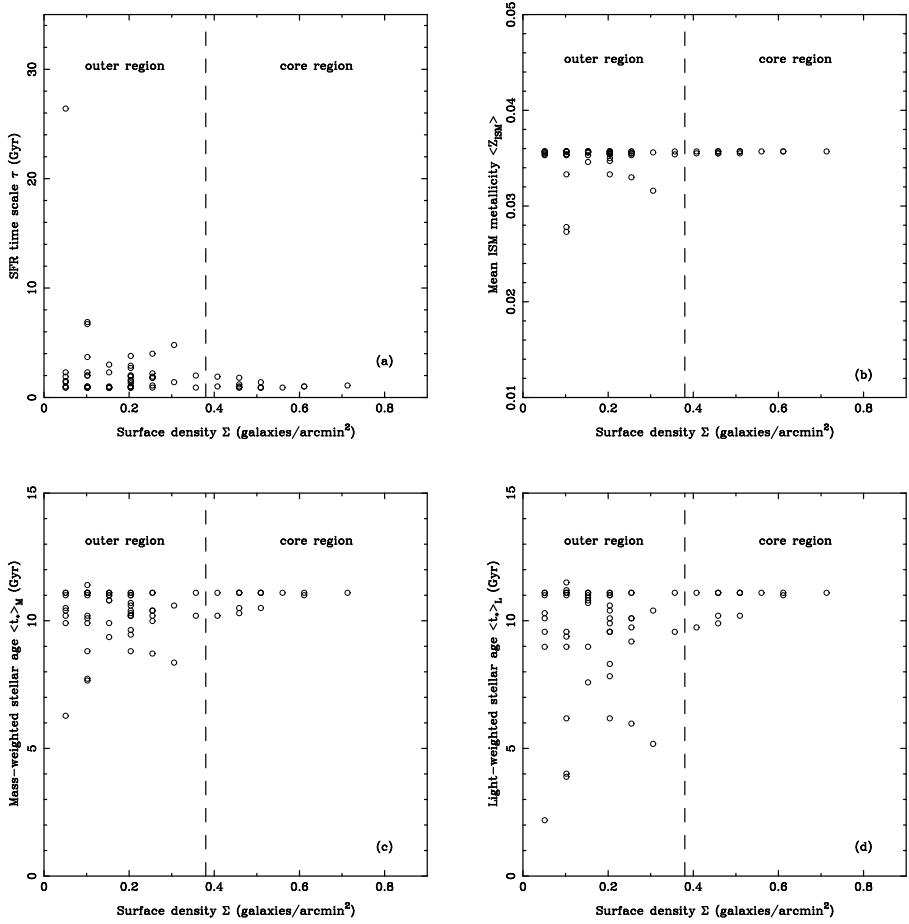}
\caption{Star formation properties for the galaxies with known
$z_{sp}$ in A98 as the functions of local surface density
$\Sigma$. The star formation properties include the SFR time scale
$\tau$, metallicity, and the mean stellar ages weighted by mass
and light.  }
\label{fig11}
\end{figure}

\begin{figure}[h]
\centering
\includegraphics[width=120mm]{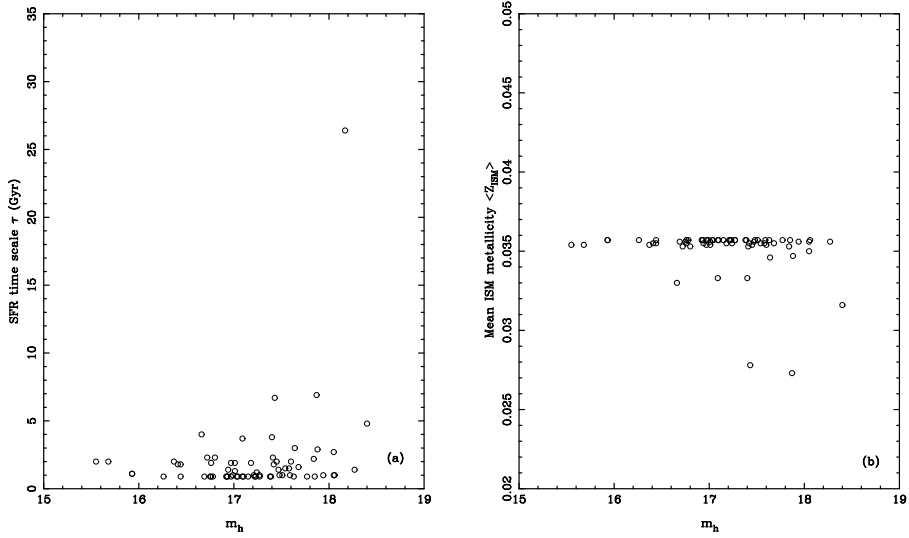}
\caption{The SFR time scales $\tau$ and mean ISM metallicities for
the galaxies with known $z_{sp}$ in A98 as the functions of
magnitude in $h$ band. }
\label{fig12}
\end{figure}

\section{Summary}

A98 is a galaxy cluster with two large enhancements in X-ray
surface brightness. This paper presents our optical photometric
observation of A98 with the Beijing-Arizona-Taiwan-Connecticut
(BATC) multicolor system. About 8,100 sources are detected down to
$V \sim 20$ mag in a field of $58' \times 58'$ centered at this
cluster, and their spectral energy distributions (SEDs) in 15
intermediate bands are obtained. There are 122 galaxies with
spectroscopic redshifts in our field, among which 74 galaxies with
$0.095 < z_{sp} < 0.115$ are selected as the members of A98. The
dynamics of two substructures (A98N and A98S) is investigated with
the help of the spectroscopic redshifts and the X-ray imaging
data. A significant substructure, A98W, is found to the 10 arcmin
west of A98S. Within our viewing field, a group of galaxies, A98X,
which contains 21 galaxies with $z_{sp} \sim 0.120$, is located
$\sim$ 15 arcmin south of A98S. According to Newtonian
gravitational binding criterion, A98X seems to be a separate
system which is gravitationally unbound to A98.

After the star-galaxy separation by the color-color diagrams, a
photometric redshift ($z_{ph}$) technique is applied to the galaxy
sample for further membership determination. The color-magnitude
relation is taken as a further restriction of the early-type
cluster galaxies. As a result, 198 galaxies with $ 0.081 < z_{ph}
< 0.113$ are selected as faint galaxy members of A98. Based on the
enlarged member galaxies, the spatial distribution, localized
velocity structure of A98 are discussed. The $\kappa$-test
algorithm supports the existing substructures A98N and A98S, and
the early suggested substructure, A98N, becomes more significant.

Assuming a Salpeter IMF and zero initial metallicity, the template
SED library with different SFRs and redshifts has been built with
the help of an evolutionary synthesis model, PEGASE. We fit the
observed SED of 74 member galaxies with known spectroscopic
redshifts one by one. The environmental effect on the star
formation histories is found for these member galaxies. The bright
massive galaxies in the core region of A98 are found to have
shorter SFR time scales, longer mean stellar ages, and higher ISM
metallicities, while the outlier galaxies are likely to have
smaller stellar ages and longer SFR time scales. This effect is
consistent with the existing correlations, such as the
morphology-density relation, the luminosity-metallicity relation,
and the mass-metallicity relation.

\normalem
\begin{acknowledgements}
We acknowledge the anonymous referee for his/her thorough reading
of this paper and invaluable suggestions. This work was funded by
the National Natural Science Foundation of China (NSFC) under
Nos.10778618 and 10633020, and by the National Basic Research
Program of China (973 Program) under No. 2007CB815403. This
research has made use of the NED, which is operated by the Jet
Propulsion Laboratory, California Institute of Technology, under
contract with the National Aeronautics and Space Administration.
We would like to thank Prof. Kong, X. at the University of Science
and Technology of China for valuable discussion.
\end{acknowledgements}

\begin{figure}[h]
\centering
\includegraphics[width=200mm]{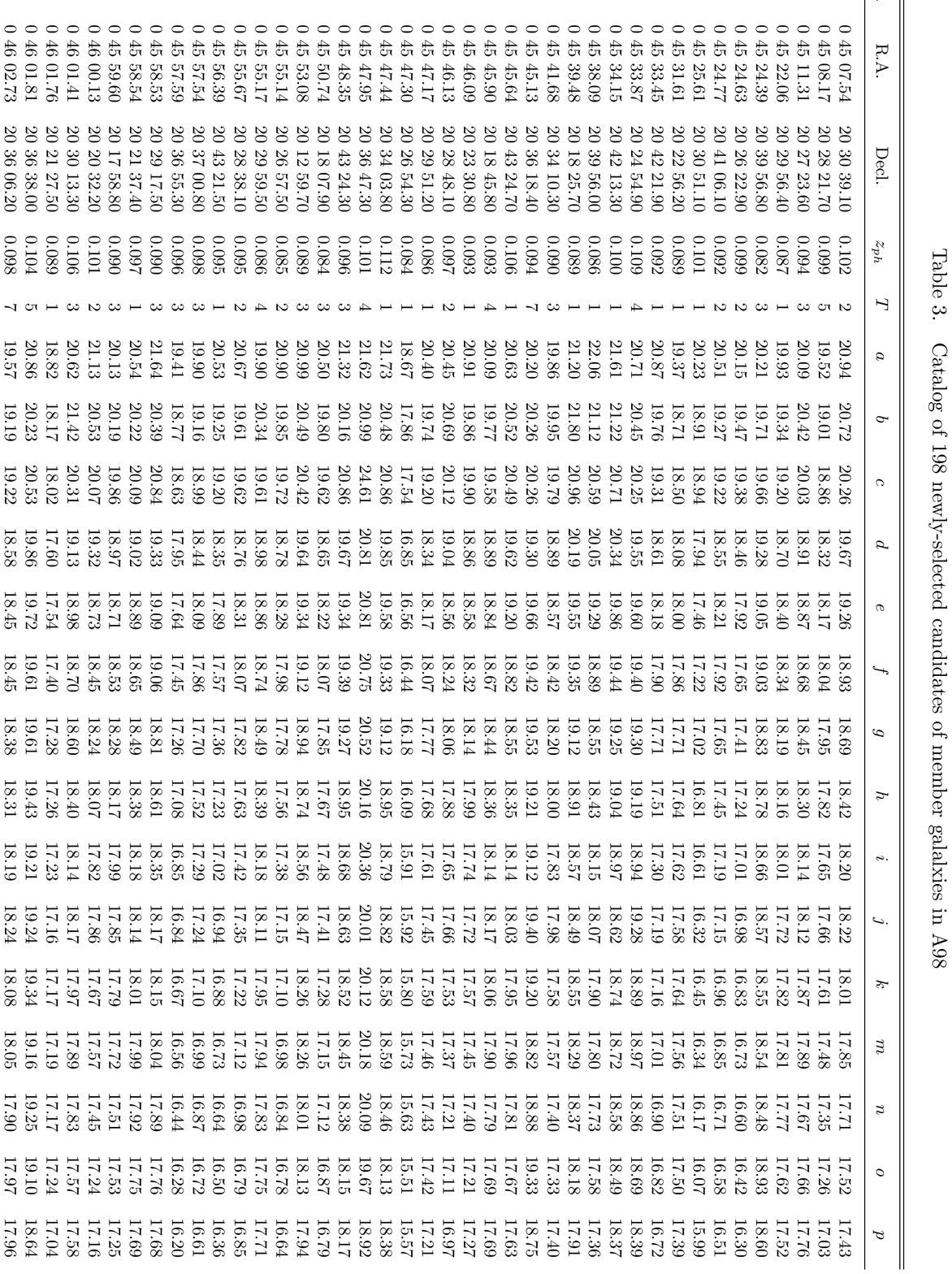}
\label{tab2_1}
\end{figure}
\begin{figure}[h]
\centering
\includegraphics[width=200mm]{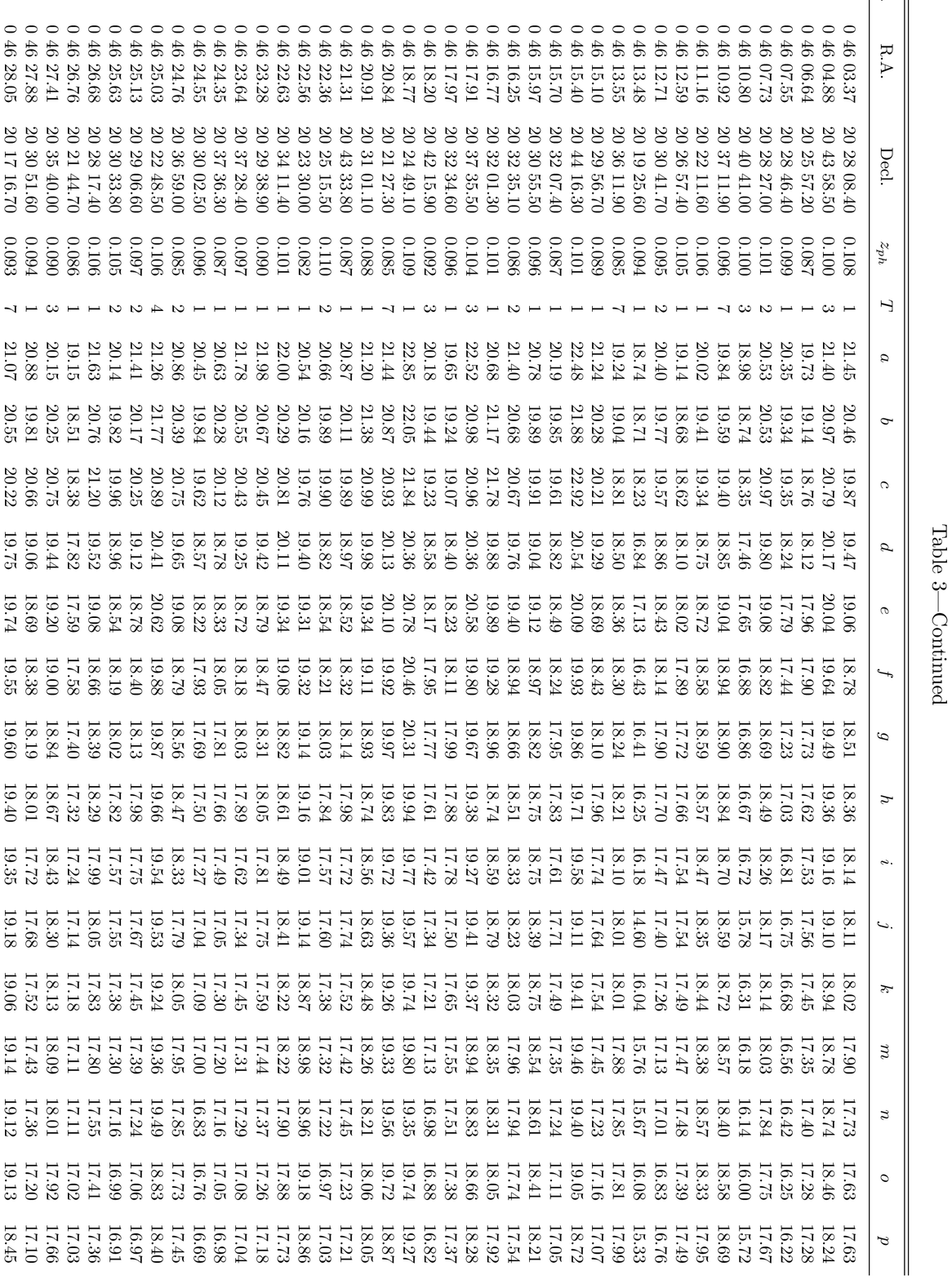}
\label{tab2_2}
\end{figure}
\begin{figure}[h]
\centering
\includegraphics[width=200mm]{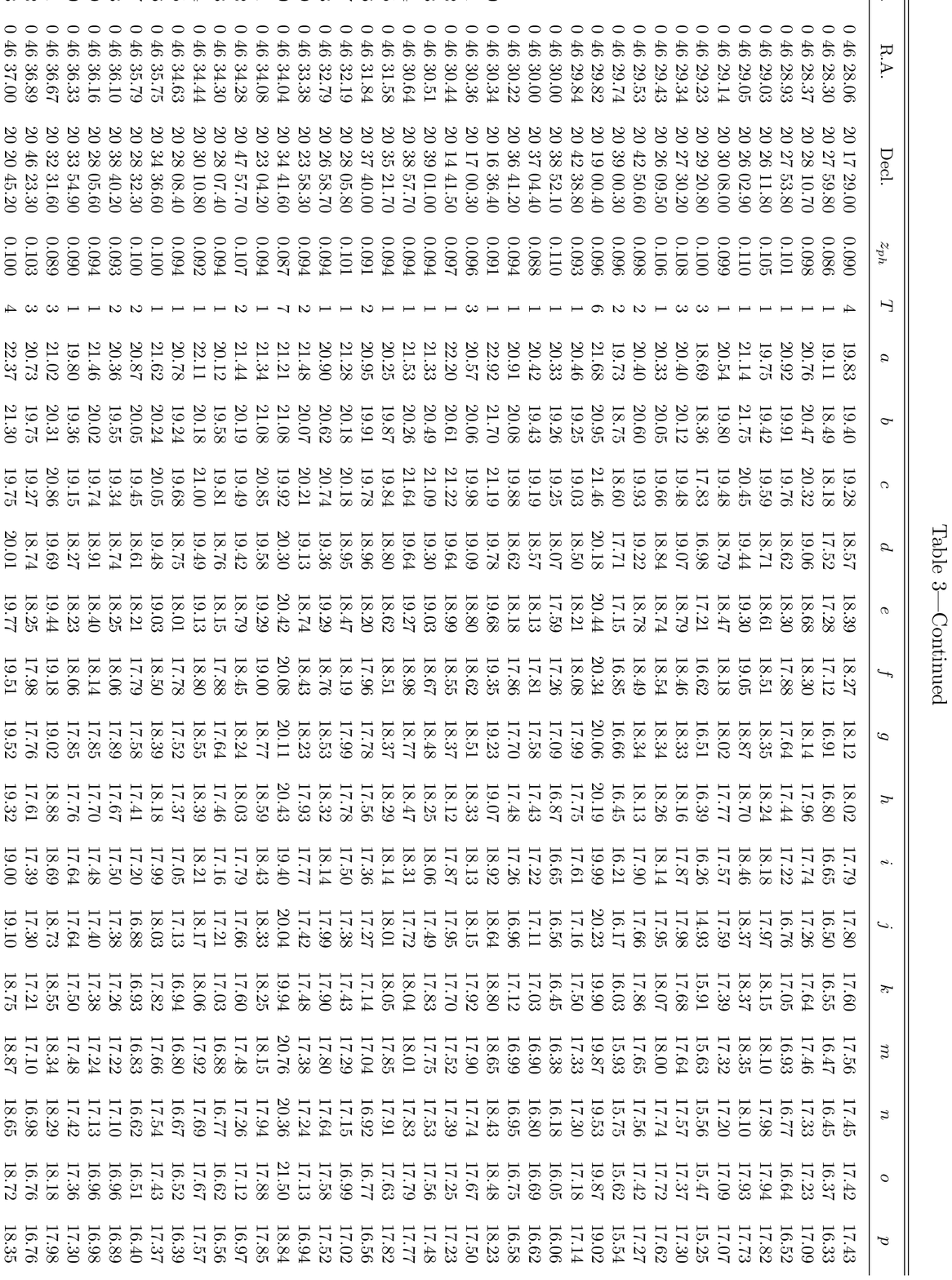}
\label{tab2_3}
\end{figure}
\begin{figure}[h]
\centering
\includegraphics[width=200mm]{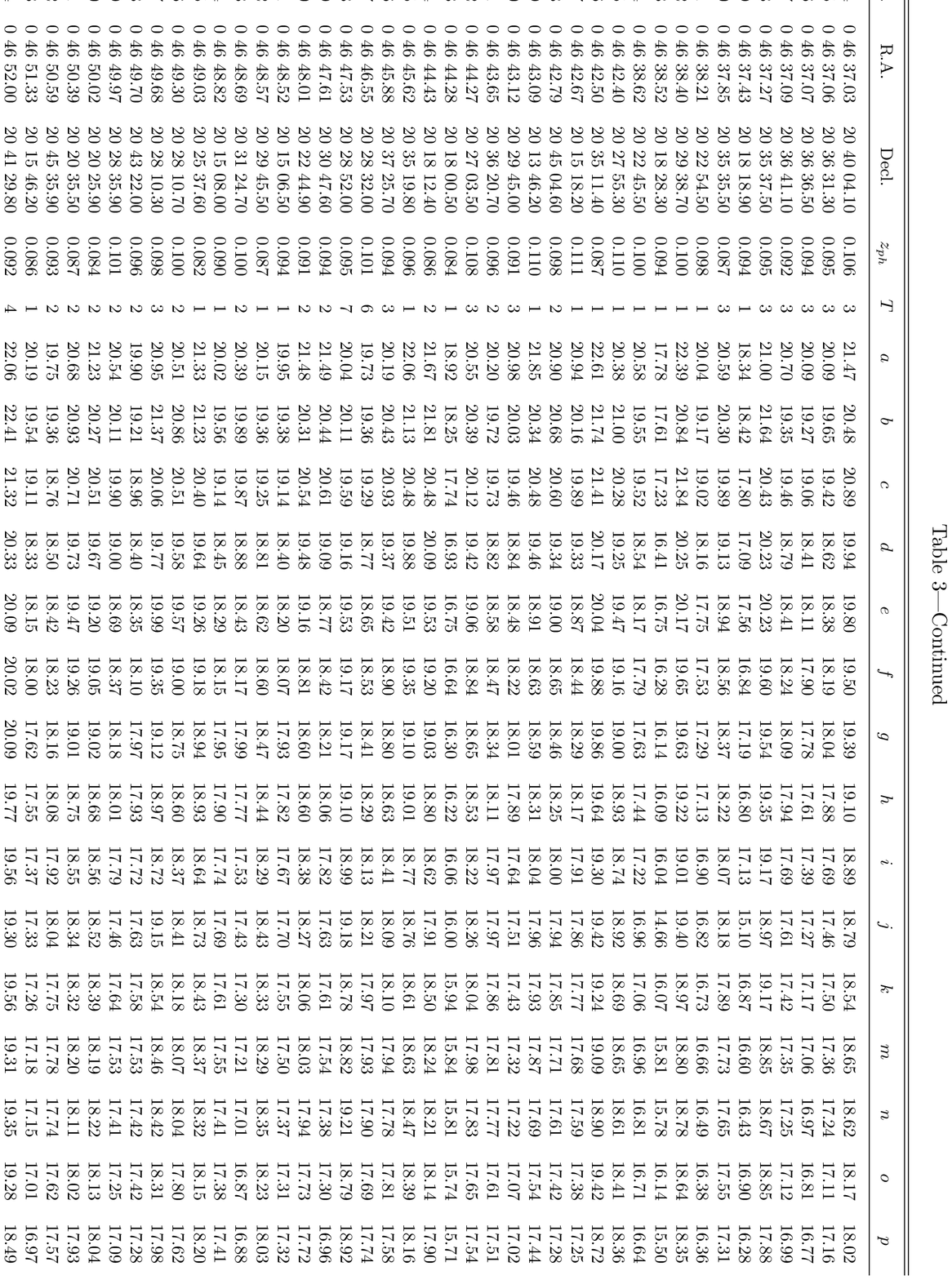}
\label{tab2_4}
\end{figure}
\begin{figure}[h]
\centering
\includegraphics[width=200mm]{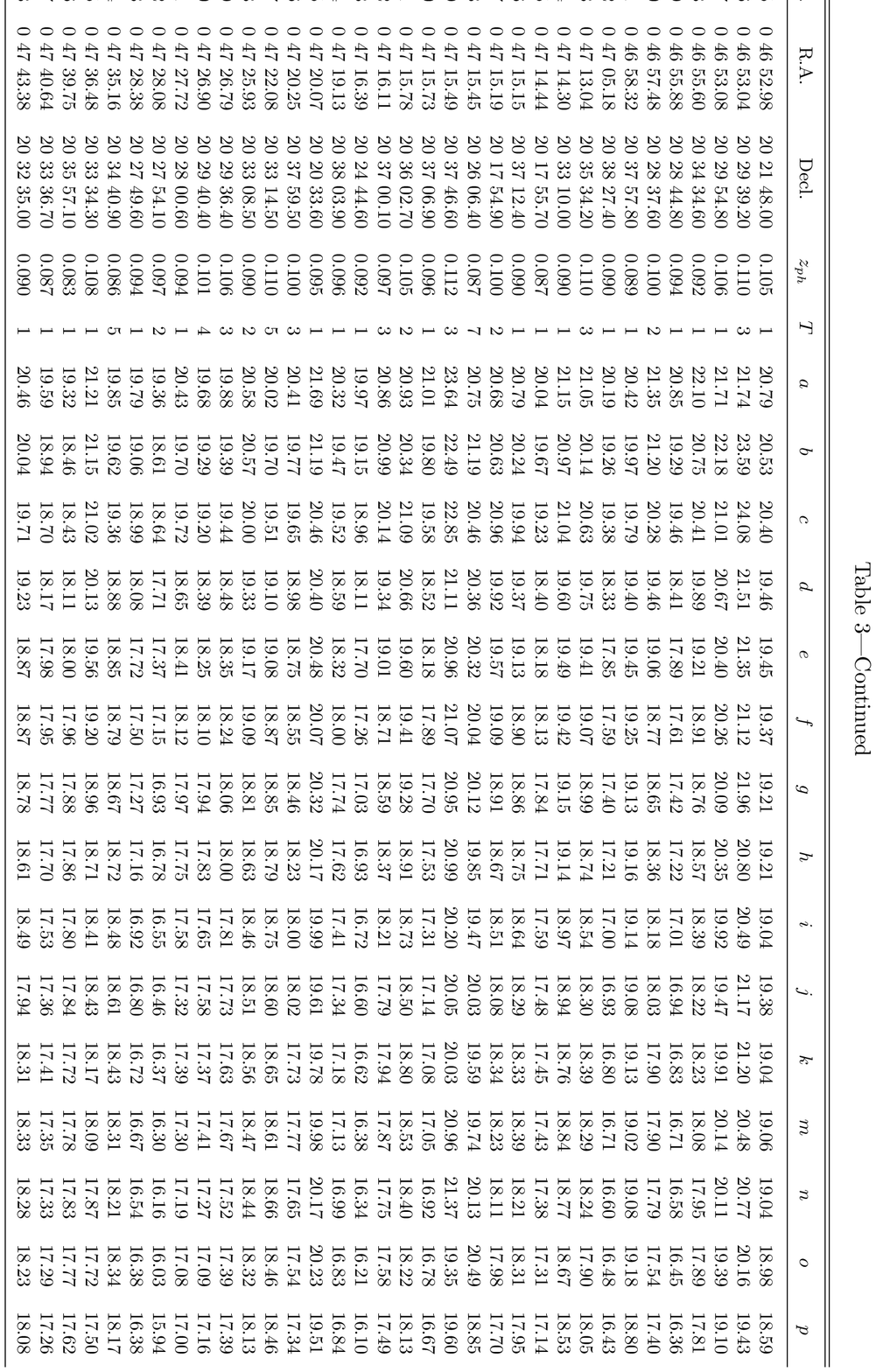}
\label{tab2_5}
\end{figure}

\label{lastpage}

\end{document}

%% file: tab1.tex
\begin{table}[t]
\bc \begin{minipage}[]{120mm}

\caption[]{Parameters of the BATC filters and the observational
statistics of A98}
\end{minipage}
\begin{tabular}{ccccccccc}   \hline
\noalign{\smallskip}
 No. & Filter & $\lambda_{c}^a$ & FWHM &
Exposure &  Number of & Seeing$^b$  & Objects & Completeness \\
   &  name & (\AA) & (\AA) &
(second) & Images     & (arcsec)  & Detected & magnitude \\
\noalign{\smallskip}   \hline \noalign{\smallskip}
  1  & a  & 3360 & 222 & 10800  & 9&  4.23   & 5107 & 21.0 \\
  2  & b  & 3890 & 291 & 3600   & 3&  3.41   & 6860 & 20.5 \\
  3  & c  & 4210 & 309 & 5400   & 6&  3.50   & 6900 & 20.5 \\
  4  & d  & 4550 & 332 & 22366  &19&  4.19   & 5753 & 20.5 \\
  5  & e  & 4920 & 374 & 12000  &10&  3.79   & 6643 & 20.5 \\
  6  & f  & 5270 & 344 & 12000  &10&  4.95   & 7977 & 20.0 \\
  7  & g  & 5795 & 289 & 7200   & 6&  4.27   & 8122 & 20.0 \\
  8  & h  & 6075 & 308 & 6000   & 5&  4.41   & 8125 & 19.5 \\
  9  & i  & 6660 & 491 & 6000   & 9&  4.05   & 8108 & 19.5 \\
  10 & j  & 7050 & 238 & 6000   & 5&  7.53   & 7503 & 19.5 \\
  11 & k  & 7490 & 192 & 7200   & 6&  3.86   & 8010 & 19.0 \\
  12 & m  & 8020 & 255 & 10800  & 9&  4.12   & 8119 & 19.0 \\
  13 & n  & 8480 & 167 & 10800  & 9&  4.38   & 7928 & 19.0 \\
  14 & o  & 9190 & 247 & 15000  &13&  3.74   & 8106 & 18.5 \\
  15 & p  & 9745 & 275 & 20400  &17&  4.93   & 7609 & 18.5 \\
\noalign{\smallskip}   \hline
\end{tabular} \ec
\tablecomments{0.86\textwidth}{ $^a$ Central wavelengths of the
filters;  $^b$ This column lists the seeings of the combined
images.}
\end{table}

%% file: tab2.tex
\begin{table}[t]
\bc \begin{minipage}[]{120mm}

\caption[]{Catalog of spectroscopically confirmed member galaxies
in A98}
\end{minipage}
\begin{tabular}{ccccc|ccccc}   \hline
\noalign{\smallskip}
 No. & R.A. & Decl & $z_{sp}$ & Ref.
 & No. & R.A.& Decl & $z_{sp}$ & Ref. \\
\noalign{\smallskip}   \hline \noalign{\smallskip}

  1 &  0 45 33.0  &  20 15 09 &  0.1003 & 1&  38  &  0  46  24.7  &  20 37 19 &  0.1028 & 2\\
  2 &  0 45 40.2  &  20 15 58 &  0.0999 & 1&  39  &  0  46  25.8  &  20 29 06 &  0.1094 & 2\\
  3 &  0 45 41.3  &  20 31 01 &  0.1030 & 1&  40  &  0  46  25.9  &  20 27 33 &  0.1041 & 3\\
  4 &  0 45 41.3  &  20 28 02 &  0.0961 & 1&  41  &  0  46  26.8  &  20 36 52 &  0.0995 & 1\\
  5 &  0 45 45.5  &  20 26 17 &  0.1045 & 2&  42  &  0  46  28.5  &  20 23 49 &  0.1013 & 4\\
  6 &  0 45 46.9  &  20 26 57 &  0.1043 & 3&  43  &  0  46  29.3  &  20 28 05 &  0.1032 & 3\\
  7 &  0 45 48.1  &  20 27 36 &  0.1034 & 2&  44  &  0  46  29.5  &  20 33 19 &  0.1045 & 3\\
  8 &  0 45 48.6  &  20 26 06 &  0.1039 & 1&  45  &  0  46  29.8  &  20 38 58 &  0.1088 & 3\\
  9 &  0 45 50.1  &  20 29 10 &  0.1012 & 1&  46  &  0  46  29.9  &  20 34 28 &  0.0978 & 1\\
 10 &  0 45 50.4  &  20 27 49 &  0.1037 & 4&  47  &  0  46  30.1  &  20 36 49 &  0.0980 & 3\\
 11 &  0 45 50.5  &  20 29 07 &  0.1033 & 1&  48  &  0  46  31.1  &  20 31 44 &  0.1034 & 1\\
 12 &  0 45 57.4  &  20 36 56 &  0.1013 & 1&  49  &  0  46  31.8  &  20 28 11 &  0.1082 & 2\\
 13 &  0 45 59.3  &  20 23 50 &  0.1058 & 1&  50  &  0  46  32.0  &  20 28 27 &  0.1048 & 1\\
 14 &  0 46 02.1  &  20 30 51 &  0.0996 & 1&  51  &  0  46  32.1  &  20 43 06 &  0.1046 & 1\\
 15 &  0 46 02.2  &  20 30 30 &  0.1060 & 1&  52  &  0  46  32.5  &  20 28 24 &  0.1060 & 1\\
 16 &  0 46 04.0  &  20 34 51 &  0.1005 & 3&  53  &  0  46  33.9  &  20 23 55 &  0.1091 & 1\\
 17 &  0 46 04.5  &  20 36 45 &  0.1044 & 4&  54  &  0  46  35.2  &  20 30 11 &  0.1041 & 1\\
 18 &  0 46 04.7  &  20 28 28 &  0.1054 & 1&  55  &  0  46  35.2  &  20 20 32 &  0.1024 & 1\\
 19 &  0 46 07.4  &  20 28 49 &  0.1059 & 1&  56  &  0  46  35.6  &  20 29 43 &  0.1067 & 3\\
 20 &  0 46 07.8  &  20 38 47 &  0.1060 & 1&  57  &  0  46  36.2  &  20 28 27 &  0.1073 & 3\\
 21 &  0 46 08.0  &  20 40 45 &  0.1019 & 2&  58  &  0  46  37.0  &  20 26 07 &  0.1023 & 4\\
 22 &  0 46 08.4  &  20 28 51 &  0.1013 & 3&  59  &  0  46  38.1  &  20 28 41 &  0.1033 & 1\\
 23 &  0 46 14.7  &  20 23 43 &  0.1034 & 1&  60  &  0  46  38.5  &  20 22 54 &  0.1037 & 4\\
 24 &  0 46 15.5  &  20 33 20 &  0.1001 & 1&  61  &  0  46  43.2  &  20 36 05 &  0.1045 & 3\\
 25 &  0 46 15.8  &  20 32 06 &  0.1024 & 1&  62  &  0  46  48.0  &  20 30 55 &  0.1036 & 1\\
 26 &  0 46 15.8  &  20 38 09 &  0.1072 & 3&  63  &  0  46  49.2  &  20 27 24 &  0.0978 & 1\\
 27 &  0 46 15.8  &  20 39 08 &  0.1075 & 2&  64  &  0  46  50.0  &  20 27 08 &  0.1055 & 1\\
 28 &  0 46 16.0  &  20 30 19 &  0.1017 & 2&  65  &  0  46  51.5  &  20 30 23 &  0.1019 & 1\\
 29 &  0 46 18.5  &  20 33 14 &  0.1051 & 1&  66  &  0  46  51.8  &  20 30 43 &  0.1062 & 1\\
 30 &  0 46 19.2  &  20 28 12 &  0.1016 & 1&  67  &  0  46  52.6  &  20 25 23 &  0.1086 & 1\\
 31 &  0 46 19.3  &  20 29 42 &  0.1098 & 2&  68  &  0  46  52.8  &  20 41 00 &  0.1051 & 3\\
 32 &  0 46 20.3  &  20 28 59 &  0.1014 & 1&  69  &  0  46  56.5  &  20 27 05 &  0.1009 & 1\\
 33 &  0 46 20.6  &  20 29 08 &  0.1032 & 1&  70  &  0  46  56.9  &  20 40 25 &  0.1031 & 2\\
 34 &  0 46 21.6  &  20 42 31 &  0.1046 & 1&  71  &  0  47  01.7  &  20 23 18 &  0.1049 & 1\\
 35 &  0 46 21.9  &  20 29 05 &  0.1116 & 3&  72  &  0  47  09.4  &  20 29 17 &  0.1051 & 1\\
 36 &  0 46 23.8  &  20 30 01 &  0.0998 & 2&  73  &  0  47  13.0  &  20 29 41 &  0.1069 & 1\\
 37 &  0 46 24.6  &  20 30 06 &  0.1043 & 3&  74  &  0  47  18.0  &  20 22 01 &  0.1049 & 1\\
\noalign{\smallskip}   \hline
\end{tabular} \ec
\tablecomments{0.86\textwidth}{ References: (1)Pinkney et al.
(2000); (2) Zabludoff et al. (1990);
 (3) Beers et al. (1982); (4) Bettoni et al. (2006).}
\end{table}